\documentclass{PoS}

\title{Constrains on the extragalactic origin of IceCube's neutrinos using HAWC}

\ShortTitle{Constrains on the extragalactic origin of IceCube's neutrinos using HAWC}

\author{Ignacio Taboada\\
        Georgia Institute of Technology \\
        E-mail: \email{itaboada@gatech.edu}}
\author{Chun Fai Tung\\
        Georgia Institute of Technology \\
        E-mail: \email{christung616@gmail.com}}
\author{\speaker{Joshua Wood}\\
       University of Wisconsin at Madison\\
       E-mail: \email{jwood@umdgrb.umd.edu}}
\author{for the HAWC Collaboration\footnote{Complete list of authors at http://www.hawc-observatory.org/collaboration/icrc2017.php}}

\abstract{
IceCube has discovered an unresolved and isotropic flux of neutrinos
between 10 TeV and 8 PeV. Extragalactic origin for this flux is
usually assumed, as well as a correlation with the sources of cosmic
rays. To date, no clear association with a class of objects has been
made. HAWC is a very-high-energy (VHE, or $\gtrsim$100~GeV) gamma ray observatory in operation in
central Mexico. 
HAWC has studied 2/3 of the entire sky ($\sim8~sr)$. We use this survey to
search for optically thin sources of cosmic rays and neutrinos, responsible for IceCube's observations. 
We
have written a simulation of the cosmological properties of neutrino
sources, FIRESONG, to show that under certain conditions, HAWC should
be observing multiple sources. 
However, HAWC has only detected 2
extragalactic objects in 760 days of livetime: Mrk 421 and Mrk 501. 
This deficiency of detected sources in the extragalactic sky can be
used to constrain the properties of neutrino sources, such as their
density in the local universe. The case HAWC restricts the most is for
no evolution in the density of sources, in which at least 8 objects
should have been detected for the cases not excluded already by IceCube.

}

\FullConference{35th International Cosmic Ray Conference – ICRC217-\\
		10-20 July, 2017\\
		Bexco, Busan, Korea}

\begin{document}

\section{Introduction}

IceCube has observed astrophysical neutrinos in the 10 TeV - 8 PeV
energy range \cite{science.1242856,PhysRevLett.115.081102}. Though
neutrino directions are known to $\sim 1^\circ$ \footnote{For the track channel,
  mostly due to CC interactions of $\nu_\mu$. Cascades, sensitive to
  all flavors, have poorer pointing.}, IceCube has not resolved
neutrino sources. In IceCube's lingo, a point source has not been found.
\cite{0004-637X-835-2-151}. 
There is not clear association of a class of objects with
astrophysical neutrinos. Multiple candidates classes have been proposed  including
Blazars, GRBs, Starburst galaxies, etc.
Isotropy is usually interpreted as indication of extragalactic origin,
however this is unproven. Finding the sources of the 
astrophysical neutrinos is one of the outstanding questions in
particle astrophysics.  

If neutrino sources are cosmic ray sources and are optically thin,
then neutrinos are due to $\pi^\pm$ decay and gamma rays are 
produced via $\pi^0$ decay. 
A common strategy to try identify the sources is to search for an
electromagnetic counterpart by pointing an instrument, from radio to
very-high-energy (VHE) gamma rays, in the direction of individual
neutrinos reported by IceCube. 
In these  
proceedings we pursue a different approach, we use 2/3 of the entire
sky that has been surveyed by the HAWC gamma ray observatory.

We will show that for a wide range of astrophysical and
cosmological options, if neutrino sources are extragalactic, HAWC should
detect multiple nearby sources. Using data collected over 760 days,
HAWC has observed 2 extragalactic sources: Mrk 421 and Mrk 501
\cite{0004-637X-843-1-40}. 
The relatively low number of observed sources can be
explained if: a) Sources are transient - currently we have only
examined time integrated HAWC data. b) The sources of neutrinos
are optically opaque. c) The neutrino spectrum has a break at
energies lower than what has been observed by IceCube. d) the
local density of neutrino sources is higher than $10^{-6}$~Mpc$^{-3}$
and sources evolve proportional to star formation history (SFH).

%In section \S\ref{sec:hawc} we describe the HAWC gamma ray
%observatory. In section \S\ref{sec:theory} current constrains by IceCube on
%the origin of the neutrino sources is described and the connexion to
%gamma rays is detailed. Section \S\ref{sec:theory} reviews
%results of analytical calculations of constrains on the density and
%luminosity of neutrino sources derived from IceCube data assuming an
%Euclidean Universe.
%Section \S\ref{sec:firesong} describes the
%FIRESONG code that we have used to simulate the extragalactic Universe
%of neutrino sources. Section \S\ref{sec:results} shows the results of
%our simulations using detailed HAWC instrument response
%information. Section\S\ref{sec:conclusions} we present a discussion of
%the results, including a comparison with observations by other
%instruments including IceCube and Fermi LAT.

\section{HAWC}
\label{sec:hawc}

HAWC is a gamma-ray observatory in operation in central Mexico at
an elevation of 4,100 m asl. HAWC operates with $>$95\% duty cycle
with a field of view of 2~sr. With Earth rotation, HAWC observes
2/3 of the sky every day. HAWC is sensitive from a few hundred GeV to
over 100 TeV gamma rays. HAWC was inaugurated on March 20, 2015, but
its modular construction allowed operation before that time. Results
presented here correspond to 760 days of effective livetime collected
between November 2014 and February 2017. Using a 507 day effective
livetime dataset, HAWC has reported the observation of 39 sources
\cite{0004-637X-843-1-40}.

HAWC uses the water Cherenkov technique, in which VHE photons are
detected by measuring Cherenkov light from ground level secondary
particles in an extensive air shower. HAWC consists of 300 steel
tanks each containing a light-tight bladder of 7.3~m diameter and
4.5~m in height. Each tank holds $\sim200,000$~liters of filtered
water. Charged particles traveling faster than the speed of light in
water produced Cherenkov radiation. A 10'' photomultiplier
tube (PMT) and three 8'' are placed at the bottom of each tank to
capture the Cherenkov light. Relative Cherenkov arrival times at each
PMT determine the direction of the shower plane, and hence the primary
gamma ray. Background is mostly cosmic ray air showers. The lateral
distribution function for cosmic rays and gamma rays air showers
differ. The former includes muons, that often land far away from the
shower core; the latter is smoother as a function of distance to the
shower core. These two facts are used
by HAWC data analysis techniques to separate signal from background.  

In these proceedings we use HAWC detailed background measured with direct
integration \cite{0004-637X-595-2-803} and the current best existing model of HAWC's
instrument response (aka HAWC pass 4.1), including the point spread
function. HAWC analysis techniques for pass 4.1 are detailed in
Ref. \cite{0004-637X-843-1-39}. 

\section{The neutrino-gamma ray connection}
\label{sec:theory}

By correlating neutrino directions and times, IceCube has ruled out
multiple source classes as the source of astrophysical  
neutrinos, including Gamma ray bursts \cite{2041-8205-805-1-L5},
Blazars \cite{icecube-blazars}, and nearby starburst galaxies \cite{0004-637X-835-2-151}.
%Gamma ray bursts (GRBs) reported by satellites can't be
%responsible of more than $\sim$0.5\% of the astrophysical neutrino
%flux \cite{2041-8205-805-1-L5}, blazars reported by LAT no more than
%17\%, the galactic plane, including both sources and diffuse
%emission, no more than 14\%, nearby starburst galaxies, no more than
%8\% . 
If neutrino sources are cosmic ray sources, then the
charged to neutral pion ratios allows the calculation of the expected
gamma ray flux \cite{gaisser-book}, with p-$\gamma$ neutrino production yielding
$\sim 2 \times$ more neutrinos than p-p interactions.  

The astrophysical neutrino spectrum is usually fitted by IceCube using a
power law. 
%Using both showers (cascades), sensitive to all
%neutrino flavors, and tracks, sensitivy to $\nu_\mu$, an
%spectral index in the range -2.1 to -2.7 has been measured. 
There may be a tension between the spectral indices
measured with tracks, $\sim\,-2.1$ \cite{PhysRevLett.115.081102}, and cascades
$\sim\,-2.7$ \cite{icecube-diffuse-icrc2015}. The discrepancy can be explained with two
neutrino populations with different spectral indices and noting that
tracks are only sensitive above $\sim$100~TeV, while cascades are
sensitive to lower energy. We do not address this issue in these
proceedings. Instead we assume a soft spectral index, since this is
the best description of IceCube's lower energy range, tens of TeV, in
which HAWC also has sensitivity. We adopt as reference the spectrum
obtained in a joint study of multiple IceCube measurements
\cite{combined} with an index of -2.5.

VHE gamma rays propagating over cosmological distances are attenuated
via the interaction with extragalactic background light (EBL). HAWC's
detectable range is $z \lesssim 0.1$ due to EBL attenuation. At
energies higher than $\sim$300 TeV, attenuation is due to interaction
with the cosmic microwave background and observable distances are of
galactic scale only. Also, HAWC's sensitivity beyond 100~TeV
(with pass 4.1) is not well understood. Hence in these proceedings we
cutoff gamma ray spectra at 100~TeV.

\section{Euclidean cosmology of neutrino sources}
\label{sec:cosmology}

\begin{figure}[tbp]
\centering
\includegraphics[width=.65\textwidth]{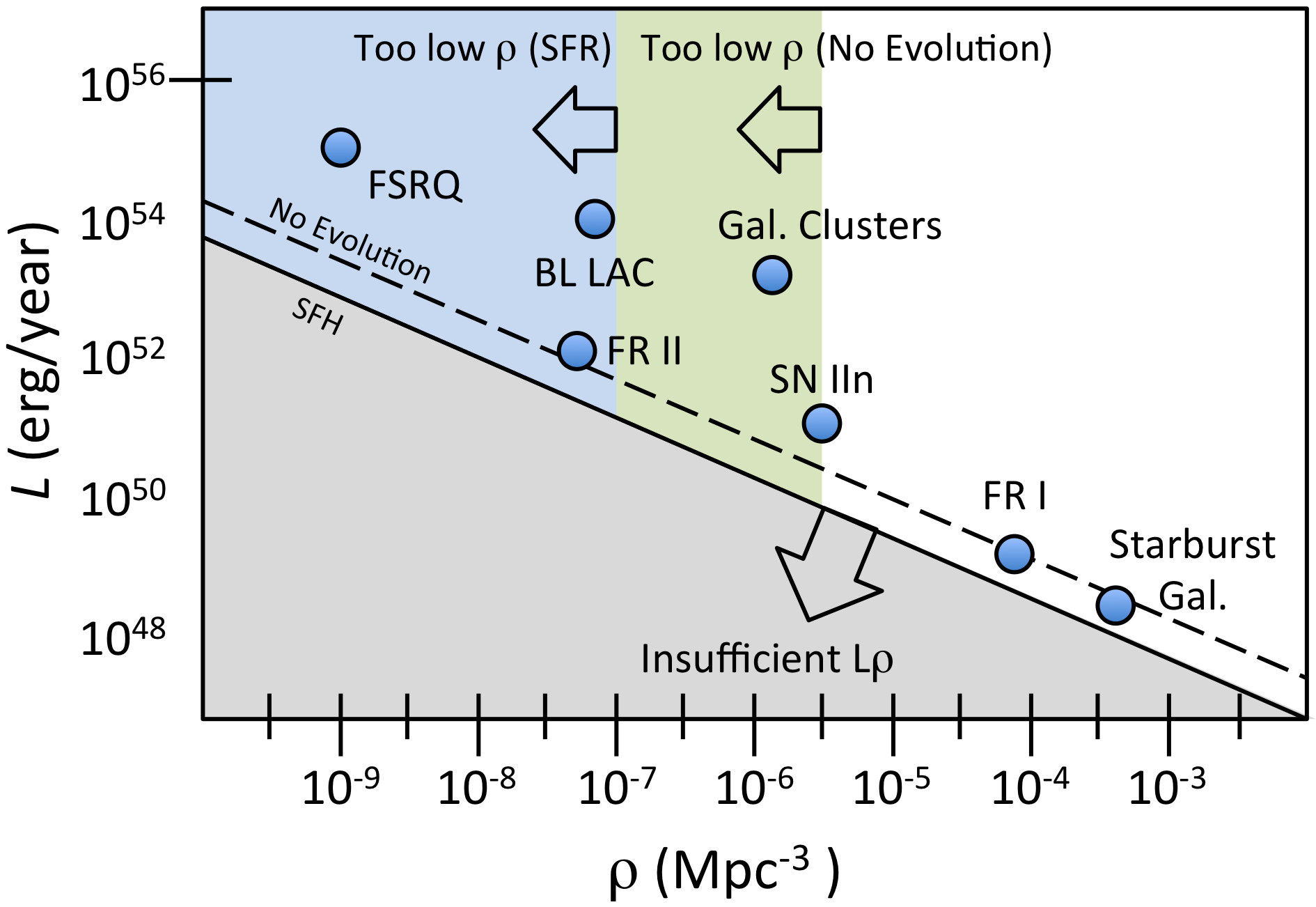}
\caption{Constrain on local power density ($\rho_0 L_\nu$)
  are shown as a diagonal lines assuming $L_\nu /L_\gamma = 1$ and
  SFH evolution for source density (solid)
  or no evolution (dashed). Also shown are lower
  bounds on local density of steady sources set by IceCube's lack of
  observation of resolved sources for SFH and no evolution cases.
}
\label{fig:kowalski}
\end{figure}

The astrophysical neutrino flux is proportional to the characteristic
luminosity of the sources in neutrinos $L_\nu$ and the density of
sources $\rho$. The density, may evolve with redshift, so
we denote $\rho_0$ as the local density. It follows that the local power density, $\rho_0 L_\nu$ is a
constant proportional to the flux measured by IceCube. Using an
Euclidean approximation, it can be shown \cite{1742-6596-632-1-012039} that:
\begin{equation}
\label{eqn:KowalskiLimit}
\rho_0 L_\nu \sim \frac{4.4\times 10^{43}}{\xi} \frac{\mathrm{erg}}{\mathrm{Mpc}^3.\mathrm{yr}},
\end{equation}
where $\xi$ is a factor in the range 0.6-8 that measures the evolution
of $\rho$. Using the simulation described in \S\ref{sec:firesong}, a
similar relationship at eqn.\ref{eqn:KowalskiLimit} is obtained. For
evolution proportional to SFH, $\xi \sim 2.6$ and
with no evolution, $\xi \sim 0.6$. SFH history evolution and no
evolution are representative of a very large number of classes of
potential sources of neutrinos such as starburst galaxies and
blazars respectively. 

The argument above assumes that sources are steady. A very
similar conclusion is reached for transient sources but with the power
density substituted by $\dot{\rho}_0 E_0$, where $\dot{\rho}$ is the
density rate of bursts and $E_0$ is the characteristic energy released
by a single burst. In these proceedings we only study the
steady case.  

A separate constrain can be derived from the lack of detection of
point sources by IceCube. The brightest (and thus nearby) sources must
have a luminosity low enough to satisfy the  lack of resolved
sources. Via the 
constrain on the power density, this lack of resolved sources implies
a lower bound on neutrino sources, $\rho \gtrsim
10^{-7}$~Mpc$^{-3} (3/\xi)^3$. Figure \ref{fig:kowalski} shows a summary of constrains
on density and power density as set by IceCube for steady sources. The
reader should note that the characteristic electromagnetic luminosity 
$L_\gamma$ may be known, but the ratio $L_\nu /
L_\gamma$ is not. Source opacity, may result in $L_\nu /
L_\gamma >1$ while non-hadronic processes would imply the opposite.
%In the VHE regime it is expected that 
%neutrino sources produce gamma rays via pion decay as well as other
%processes, e.g. inverse Compton scattering, so it'd would be expected
%that $L_\nu /L_\gamma<1$. Then again, if the sources are opaque, even
%modestly opaque, the inequality can be reversed. 

\section{FIRESONG: a simulation of neutrino source cosmology}
\label{sec:firesong}

\begin{figure}[tbp]
\centering
\includegraphics[width=.5\textwidth]{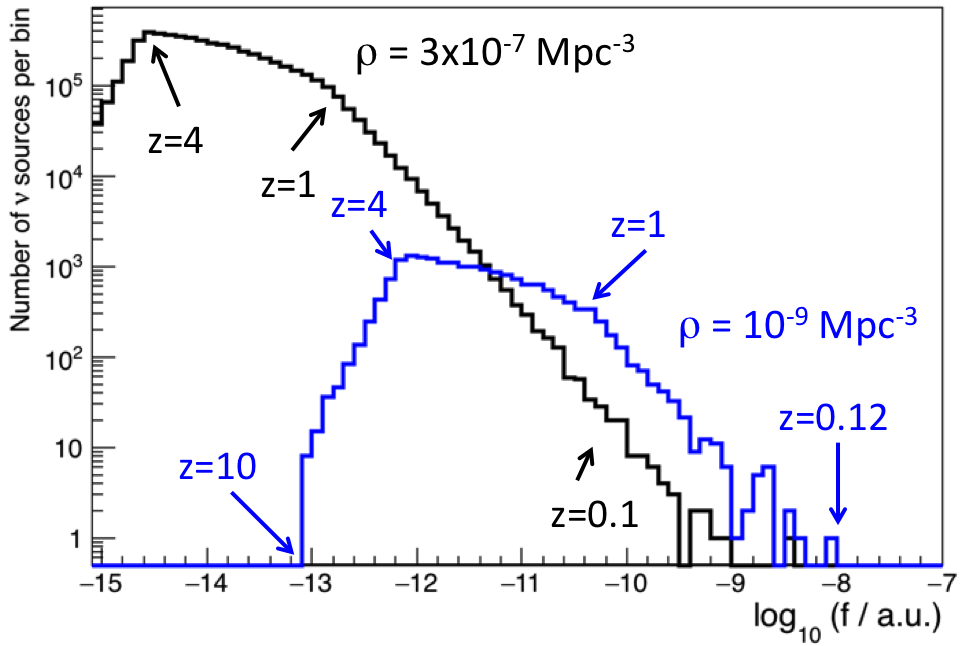}
\caption{Sample FIRESONG output for two choices of local density of
  neutrino sources. These two simulation runs assumed SFH 
  evolution and standard candle luminosity function.}
\label{fig:FIRESONG}
\end{figure}

We have written a simulation of neutrino sources, FIRESONG, that replicates
\S\ref{sec:cosmology}, but using $\Lambda$CDM cosmology. The total
flux of all neutrino sources simulated is made to match IceCube's
observations, e.g. Ref. \cite{combined}.
FIRESONG
includes many choices of source evolution, such as SFH as determined by Hopkins and Beacom \cite{0004-637X-651-1-142} and
Candels and Clash for evolution following core collapse supernovae\cite{0004-637X-813-2-93}. FIRESONG also
allows for choice of luminosity functions, such as power laws,
lognormal distributions, standard candles, etc. The code is very
modular and easily allows coding of additional density evolution and
luminosity functions, including evolution of the luminosity
function. Though in these proceedings we only deal with steady
sources, FIRESONG can also simulate transient sources. FIRESONG can be
used in two modalities. In the mode used in 
the current proceedings the local density of sources is a free
parameter. Once this has been set, the output is the list of all
neutrino sources in the universe, described by their neutrino flux at
Earth and redshift. A separate mode, not used here, produces
individual neutrino events, also described by neutrino flux and
redshift. This latter mode can be used by observatories that point in
the direction of individual neutrinos to try to find an
electromagnetic counterpart. FIRESONG code and documentation can be
downloaded from \cite{firesong}. 

A sample output of FIRESONG can be seen in
fig. \ref{fig:FIRESONG}. The panels show the number of sources for a
given neutrino flux at Earth for two choices of local density. The
simulation was performed assuming SFH evolution \cite{0004-637X-651-1-142} and with standard candle sources. This
choice of luminosity function means a direct mapping between flux at
Earth and redshift. Various redshifts are indicated in the
figures. The highest redshift considered is 10. At these redshift,
evolution of density and luminosity is very uncertain, however, the
bulk of the contribution to the neutrino flux observed at Earth is for
$z \lesssim 2$.

Fig. \ref{fig:FIRESONG} shows that, as expected, higher local density
means more sources in the Universe. But also as described in section
\S\ref{sec:cosmology}, higher density results in lower luminosity
(flux) sources. For a 
density of $3\times 10^{-7}$~Mpc$^{-3}$ multiple sources in the
Universe are at a redshift of 0.1 or lower, potentially detectable by
HAWC. But for density $10^{-9}$~Mpc$^{-3}$ the nearest source
in this simulation run had a redshift of 0.12 - possibly at the edge
of what HAWC can detect.

\section{All sky sensitivity of HAWC to neutrino sources}
\label{sec:results}

\begin{figure}[tbp]
\centering
\includegraphics[width=.45\textwidth]{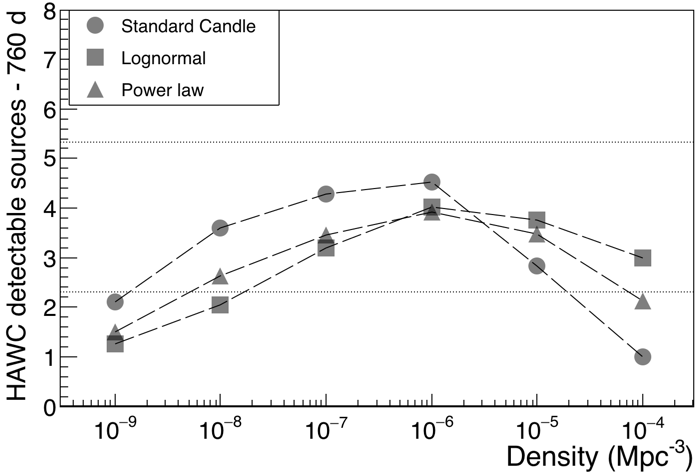}
\includegraphics[width=.45\textwidth]{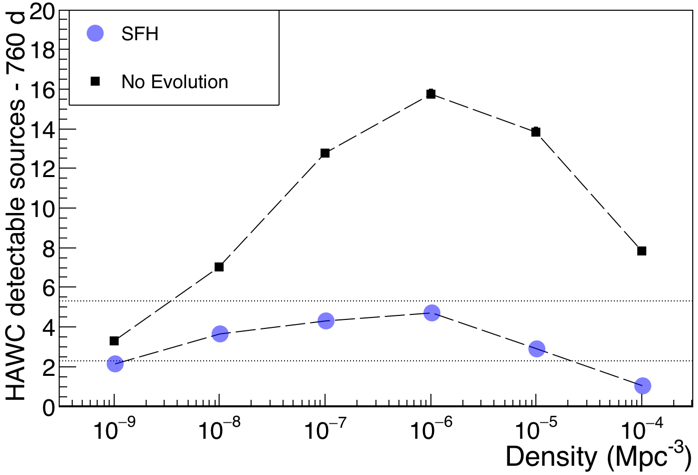}
\caption{Number of sources detectable by HAWC in 760 days of
  operation. Left pannel corresponds source density that evolves
  following SFH \cite{0004-637X-651-1-142}. Various choices of luminosity functions
  are shown. The right panel compares standard candle
  sources with evolution as SFH and no evolution. }
\label{fig:Detectable}
\end{figure}

To convert neutrino fluxes produced by FIRESONG we assume p-p neutrino
production because this is the more pessimistic scenario for gamma
rays. We also apply EBL attenuation following Gilmore et 
al. \cite{doi:10.1111/j.1365-2966.2009.15392.x}. Source location in
the sky, $\delta$ and RA, are randomize assuming isotropy. Each
FIRESONG run  
results in a list of nearby sources described by and EBL attenuated
gamma ray spectrum and a location in the sky. 

We use HAWC's software to inject the sources from each FIRESONG run
into a background template map measured by HAWC using direct
integration. Then the signal strength (in significance) is
determined. The determination of the significance of a source uses
HAWC's knowledge of the point spread function for gamma rays in terms
of, e.g. declination and gamma ray energy \footnote{Event size
  actually, which is correlated with primary energy.}. In HAWC, to achieve a post-trial significance
over the entire sky above 5 sigma is approximately equivalent to a
pre-trial significance of 7.5 sigma. We consider a FIRESONG simulated
source detectable if it exceeds 7.5 sigma pre-trial using HAWC's software. 

Figure \ref{fig:FIRESONG} provides guidance into the range of local
densities HAWC can explore. At very low densities, even the nearest
sources are too far to be detectable by HAWC due to EBL
attenuation. Using \S\ref{sec:cosmology} it can be shown that the
flux of the nearest sources depends on $\rho^{-1/3}$. As the density
increases, the nearest sources have a redshift that is more favorable
for HAWC but the flux only decreases slowly. When the density is very
high, there's little or no EBL attenuation at all (i.e. when there
are multiple nearby sources at a few tens of Mpc). Beyond this
density, the number of HAWC detectable sources diminishes as sources
become weaker and weaker with increasing density. In summary, we 
expect HAWC to be most sensitive to intermediate values, $\rho \;
10^{-7} \,-\, 10^{-5}$~Mpc$^{-3}$ of local density of neutrino sources. 

\begin{figure}[tbp]
\centering
\includegraphics[width=.45\textwidth]{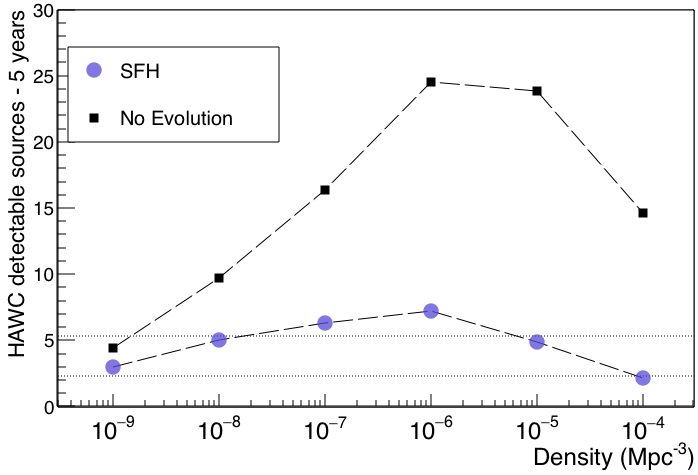}
\caption{Predicted number of sources detectable by HAWC in 5 years of
  operation. Standard candles following SFH as well as no evolution.}
\label{fig:Detectable5yr}
\end{figure}

Figure \ref{fig:Detectable} shows the number of FIRESONG predicted
detectable sources by HAWC in 760 days of operation. The left panel
shows density evolving as SFH for various luminosity
functions: Standard candles, a power
law luminosity function $dN/dL \propto L^{-2}$ over 3 orders of
magnitude in the range of luminosity and a
lognormal function with a width of proportional to the average
luminosity, $L_\nu$.
The various
choices of luminosity function only result in a factor of $\sim$2 in
the number of detectable sources, expect at the 
highest density considered.  However, a factor of $\sim 3-5$ results
from a change from SFH evolution to no evolution (right panel
fig. \ref{fig:Detectable}). The reason why the no evolution is more
favorable for HAWC is that the number of nearby sources does not
change much. Within redshift 0.1 is $\sim 23$ for SFH evolution and $\sim 17$ for no
evolution. But for no evolution, the luminosity is $\sim$4 times
larger than for SHF, as expected from
eqn. \ref{eqn:KowalskiLimit}. Figure \ref{fig:Detectable5yr} shows
standard candles for 5 years of HAWC operation with SFH
and no evolution. 

As a guideline of whether these calculations are inconsistent or not
with HAWC observations, figures \ref{fig:Detectable} and \ref{fig:Detectable5yr} show
horizontal dashed lines at 2.3 (5.3) detectable sources. These are the
numbers at which a given prediction is inconsistent at 90\% C.L. with
an observation of zero (two) sources. 

% Why is no evolution better for HAWC?
%
% Test rho = 1e-7 Mpc^-3, SFH, Standard Candles, index 2.5
% Luminosity: 5.7082e+50 erg/yr
% 35 sources closer than z = 0.1 - or 23 in HAWC's sky
%
% Test rho = 1e-7 Mpc^-3, No Evolution, Standard Candles, index 2.5
% Luminosity: 2.1805e+51 erg/yr
% 25 sources closer than z = 0.1 - or 17 in HAWC's sky
% 
%
%

\section{Discussion}
\label{sec:conclusions}

HAWC has observed two sources in the extragalactic sky. Are the
results presented here consistent of not with that observation? The
case of Mrk 421 and 501 being cosmic ray (and neutrino) sources is not
settled. Both can be described with purely leptonic or with a mixed
leptonic/hadronic model. 

Let's assume that they are sources of neutrinos. 
Blazars, such as Mrk 421 and 501 do not have evolution in their
density.  As we found in \S\ref{sec:results}, a wide range of densities is disfavored by HAWC
for no evolution. Blazars, however, have evolution in their
luminosity. We have not simulated this case in detail. Murase and Waxman,
\cite{PhysRevD.94.103006} define an ``effective''  local density for
sources to take into account luminosity evolution. 
%For Blazars their
%local density is $\rho_0\sim 10^{-7}$~Mpc$^{-3}$, but their effective 
%density is $\rho_0^{eff} \sim 10^{-8}$~Mpc$^{-3}$. Using
%$\rho_0^{eff}$, our results are inconsistent with Blazars, including
%Mrk 421 and 501 being the sources of astrophysical neutrinos.
%This conclusion will be studied in more detail in future work.
We will explore the case of blazars in more detail in future work.

Let's asume now that Mrk 421 and 501 are not neutrino sources. A broad
range of densities, from $10^{-8}$ to $10^{-5}$~Mpc$^{-3}$,
are disfavored, regardless of density evolution.
 
What characteristics the neutrino sources must have to avoid the
constrains presented here: a) they can be transient - which we haven't
studied yet b) the neutrino sources may be opaque to gamma rays c)
there is a break in the neutrino spectrum at $\sim$10~TeV.

Our conclusions support previous work \cite{0004-637X-836-1-47} that
found that the astrophysical neutrino flux is too high in comparison
with Fermi's extragalactic gamma ray background. Indeed Kistler
\cite{kistler} proposed the break at 10~TeV (and matching p-$\gamma$
neutrino production) to solve the incompatibility. At the highest
densities, $\sim 10^{-5}$~Mpc$^{-3}$ HAWC is sensitive to gamma rays
in the tens of TeV energy range, because sources are little or not
affected by EBL. Hence at the highest densities, a break in the
spectrum may not provide an answer to the apparent excess of neutrinos
in comparison to gamma rays.

\section*{Acknowledgements}

HAWC Collaboration: http://www.hawc-observatory.org/collaboration/icrc2017.php

%\begin{thebibliography}{99}
\bibliographystyle{abbrv}

\end{document}